\documentclass[pre,aps,preprint,showpacs,preprintnumbers,amsmath,amssymb,secnumarabic,eqsecnum,eqappnum]{revtex4}

\usepackage{graphicx}
\usepackage{dcolumn}
\usepackage{bm}

\newcommand{\beq}{\begin{equation}}
\newcommand{\eeq}{\end{equation}}
\newcommand{\beqa}{\begin{eqnarray}}
\newcommand{\eeqa}{\end{eqnarray}}
\newcommand{\vc}[1]{\mbox{\boldmath $#1$}}

\newcommand{\vol}[1]{{\bf #1}}

\newcommand{\du}[1]{{\bf\sf #1}}

\begin{document}


\title{Swimming of a uniform deformable sphere in a viscous incompressible fluid with inertia}

\author{B. U. Felderhof}

 \email{ufelder@physik.rwth-aachen.de}
\affiliation{Institut f\"ur Theorie der Statistischen Physik\\ RWTH Aachen University\\
Templergraben 55\\52056 Aachen\\ Germany\\
}%

\author{R. B. Jones}

 \email{r.b.jones@qmul.ac.uk}
\affiliation{Queen Mary University of London, The School of
Physics and Astronomy, Mile End Road, London E1 4NS, UK\\}%

\date{\today}

\begin{abstract}
The swimming of a deformable uniform sphere is studied in second order perturbation theory in the amplitude of the stroke. The effect of the first order reaction force on the first order center of mass velocity is calculated in linear response theory by use of Newton's equation of motion. The response is characterized by a dipolar admittance, which is shown to be proportional to the translational admittance. As a consequence the mean swimming velocity, calculated in second order perturbation theory, depends on the added mass of the sphere. The mean swimming velocity and the mean rate of dissipation are calculated for several selected strokes.
\end{abstract}

\maketitle
\section{\label{1}Introduction}

The swimming of a deformable body in an incompressible fluid poses a challenging problem of fluid mechanics. At low Reynolds number the fluid motion is laminar and the periodic shape deformation of the body causes a mean uniform velocity of its center of mass, even though the mean force on the body vanishes. In general the time-dependent swimming velocity has an oscillatory component due to the periodic reaction force generated by the dipolar moment of the shape deformation. In the following the phenomenon is analyzed on the basis of explicit calculations for a deformable sphere.

In recent work \cite{1} on the dynamics of swimming of a deformable sphere in a viscous incompressible fluid with inertia we showed that if the distortion of the spherical surface has an oscillatory dipolar component, then a reaction force is generated which can cause an oscillatory motion of the sphere. To first order the motion is linear in the amplitude of the dipolar distortion. Its magnitude depends on the inner dynamics of the swimmer. In earlier work on swimming in a fluid with inertia \cite{2} we put the first order velocity equal to zero. We view this now as a kinematic condition which is realized only if the reaction force is fully absorbed by the sphere and has no effect on its surface motion. More generally, the effect of the oscillatory reaction force must be taken into account.

For a uniformly deforming sphere the velocity of the surface is identical with that of its center of mass, and the linear velocity cannot be neglected. In the following we show that the linear oscillatory motion affects the first order flow pattern and hence also the mean swimming velocity, where the mean is defined as the time average over a period of the stroke. The effect is dependent on the mass density of the sphere.

The effect of fluid inertia on swimming is characterized by a scale number $s$, defined by $s=a\sqrt{\omega\rho/2\eta}$, where $a$ is the radius of the undistorted sphere, $\omega$ is the frequency of the stroke, $\rho$ is the mass density of the fluid, and $\eta$ is its shear viscosity. In the Stokes limit $s=0$ inertia can be neglected. In the limit of large $s$ the swimming is dominated by fluid inertia.

We showed earlier \cite{3} for situations with vanishing linear velocity that the mean swimming velocity and the mean rate of dissipation depend intricately on the scale number $s$. For certain strokes the mean swimming velocity can change sign as $s$ increases. Such a reversal of swimming velocity was found also in computer simulation of a two-sphere system by Jones et al. \cite{4}, Dombrowski et al. \cite{5}, and Dombrowski and Klotsa \cite{6}. In their model the two spheres are rigid, but oscillate relative to each other. The model was introduced earlier by Klotsa et al. \cite{7}. A mechanical model with simplified hydrodynamic interactions \cite{8} does not show the reversal of mean swimming velocity. Apparently the reversal is related to the details of the oscillatory flow pattern and the combination of friction and inertia.

The effect of fluid inertia on the swimming of a deformable sphere was first studied by Rao \cite{9}. The effect of Reynolds stress on the motion of a squirmer was studied by Wang and Ardekani \cite{10}, by Khair and Chisholm \cite{11}, and by Chisholm et al. \cite{12}. Spelman and Lauga \cite{13} studied the translational velocity of a squirmer in the inertia-dominated limit by the method of matched asymptotic expansion. Inertial effects on the hydrodynamic interaction of two swimmers were studied numerically by Li et al. \cite{14}.

In Sec. 3 of this article we elucidate the mechanism which couples the dipolar surface distortion to the linear motion of a uniform sphere. The reaction force which generates the linear center of mass motion must be calculated self-consistently. The linear response to the dipolar distortion follows from Newton's equation and is characterized by a transport coefficient, which we call the dipolar admittance. This quantity depends on the scale number $s$ and the ratio of mass densities $\rho_0/\rho$, and turns out to be proportional to the translational admittance.

In Secs. 4 and 5 we calculate elements of the two matrices which enter the calculation of the mean swimming velocity and the mean rate of dissipation in second order perturbation theory in the amplitude of the stroke. As before \cite{3} it is useful to introduce a Stokes representation to facilitate the calculation. In Sec. 6 we study the mean swimming velocity and the mean rate of dissipation as functions of scale number and the ratio of mass densities for a number of different strokes. The article is concluded with a discussion.

\section{\label{2}Fluid motion}

We consider a flexible sphere of radius $a$ immersed in a viscous
incompressible fluid of shear viscosity $\eta$ and mass density $\rho$.
The fluid is set in motion by time-dependent distortions of the
sphere. We shall study axisymmetric periodic distortions which lead to a translational swimming
motion of the sphere in the $z$ direction in a Cartesian system of coordinates. The analysis is based on a perturbation expansion of the Navier-Stokes equations in powers of the amplitude of distortions \cite{2}. The no-slip boundary condition is applied on the surface of the distorted sphere.

The surface distortion is written as
\begin{equation}
\label{2.1}
\vc{\xi}(\theta,t)=\mathrm{Re}[\vc{\xi}_\omega(\theta)e^{-i\omega t}],
\end{equation}
with polar angle $\theta$ and complex amplitude $\vc{\xi}_\omega(\theta)$. The distortion $\vc{\xi}$ is defined in the co-moving frame of the body with the origin at rest. Below we consider in particular distortions with the property $\vc{\xi}_\omega(\theta)\cdot\vc{e}_\varphi=0$, where $\vc{e}_\varphi$ is the unit vector in the azimuthal direction. In our recent analysis of the dipole-quadrupole (DQ) model and the quadrupole-octupole (QO) model \cite{15} we presented graphical displays of this type of distortions.

We showed that in the DQ model the generated flow gives rise to a first order oscillating force on the body which in general leads to a first order motion \cite{1}. By definition we put the first order velocity equal to zero, in accordance with our earlier considerations \cite{2}, and we argued that this can be justified on the basis of an assumption on the inner structure of the body \cite{1}.

More generally we must allow internal dynamics for which the first order velocity does not vanish. In this article we assume in particular that the distorting sphere remains uniform with mass density $\rho_0$. If the distortion has a dipolar component then the hydrodynamic reaction force on the body leads to an oscillating first order velocity $\vc{U}_1(t)=U_1(t)\vc{e}_z$ with vanishing mean.

The flow velocity $\vc{u}(\vc{r},t)$ and pressure $p(\vc{r},t)$ in the laboratory frame are assumed to satisfy the Navier-Stokes equations
\begin{equation}
\label{2.2}\rho\bigg[\frac{\partial\vc{u}}{\partial t}+\vc{u}\cdot\nabla\vc{u}\bigg]=\eta\nabla^2\vc{u}-\nabla p,\qquad\nabla\cdot\vc{u}=0.
\end{equation}
We consider a solution of these equations which varies periodically, as caused by the periodically varying shape of the body. In the laboratory frame the flow velocity tends to zero at infinity and the pressure tends to the ambient value $p_0$. The periodicity of the solution implies
\begin{eqnarray}
\label{2.3}\vc{u}(\vc{r}-\overline{U}T\vc{e}_z,t+T)&=&\vc{u}(\vc{r},t),\nonumber\\
p(\vc{r}-\overline{U}T\vc{e}_z,t+T)&=&p(\vc{r},t),
\end{eqnarray}
where $\overline{U}$ is the mean swimming velocity, $T=2\pi/\omega$ is the period, and $\vc{e}_z$ is the unit vector in the $z$ direction.
The mean swimming velocity $\overline{U}$ is of second order in the surface distortion $\vc{\xi}$.

We can assume that at time $t=0$ the centroid of the body is at the origin. To  first order after period $T$ the centroid is again at the origin. The first order velocity and pressure take the form
\begin{equation}
\label{2.4}\vc{u}^{(1)}(\vc{r},t)=\mathrm{Re}[\vc{u}_\omega(\vc{r})e^{-i\omega t}],\qquad p^{(1)}(\vc{r},t)=\mathrm{Re}[p_\omega(\vc{r})e^{-i\omega t}]
\end{equation}
with amplitudes $\vc{u}_\omega(\vc{r}),p_\omega(\vc{r})$
which satisfy the linearized Navier-Stokes equations
\begin{equation}
\label{2.5}\eta[\nabla^2\vc{u}_\omega-\alpha^2\vc{u}_\omega]-\nabla p_\omega=0,\qquad\nabla\cdot\vc{u}_\omega=0,
\end{equation}
with the variable
\begin{equation}
\label{2.6}\alpha=(-i\omega\rho/\eta)^{1/2}=(1-i)(\omega\rho/2\eta)^{1/2}.
\end{equation}
The solution of Eq. (2.5) can be expressed as a linear superposition of modes \cite{16}
 \begin{eqnarray}
\label{2.7}\vc{v}_l(\vc{r},\alpha)&=&\frac{2}{\pi}\;e^{\alpha a}[(l+1)k_{l-1}(\alpha r)\vc{A}_l(\hat{\vc{r}})+lk_{l+1}(\alpha r)\vc{B}_l(\hat{\vc{r}})],\nonumber\\
\vc{u}_l(\vc{r})&=&-\bigg(\frac{a}{r}\bigg)^{l+2}\vc{B}_l(\hat{\vc{r}}),\qquad p_l(\vc{r},\alpha)=\eta\alpha^2a\bigg(\frac{a}{r}\bigg)^{l+1}P_l(\cos\theta),
\end{eqnarray}
with modified spherical Bessel functions $k_l(z)$ \cite{17}, radial unit vector $\hat{\vc{r}}=\vc{r}/r=\vc{e}_r$, and vector spherical harmonics $\{\vc{A}_l,\vc{B}_l\}$ defined by \cite{16}
 \begin{eqnarray}
\label{2.8}\vc{A}_l&=&\hat{\vc{A}}_{l0}=lP_l(\cos\theta)\vc{e}_r-P^1_l(\cos\theta)\vc{e}_\theta,\nonumber\\
\vc{B}_l&=&\hat{\vc{B}}_{l0}=-(l+1)P_l(\cos\theta)\vc{e}_r-P^1_l(\cos\theta)\vc{e}_\theta,
\end{eqnarray}
with Legendre polynomials $P_l$ and associated Legendre functions $P^1_l$ in the notation of Edmonds \cite{18}.

The surface distortion function $\vc{\xi}_\omega(\vc{s})$  in the co-moving frame is prescribed and expanded as
\begin{equation}
\label{2.9}\vc{\xi}_\omega(\vc{s})=-ia\sum^\infty_{l=1}[\kappa_l\vc{v}_l(\vc{s},\alpha)+\mu_l\vc{u}_l(\vc{s})],
\end{equation}
with $\vc{s}=a\hat{\vc{r}}$ and complex coefficients $\{\kappa_l,\mu_l\}$. The mode $\vc{v}_1(\vc{s},\alpha)$ involves the vector spherical harmonic $\vc{A}_1=\vc{e}_z$ corresponding to uniform displacement. The absence of uniform displacement in the co-moving frame implies the constraint $\kappa_1=0$. The first order fluid velocity at the surface in the co-moving frame is given by $\vc{v}^{(1)}(\vc{s},t)=\partial\vc{\xi}(\vc{s},t)/\partial t$ according to the no-slip boundary condition.
In the laboratory frame the surface is assumed to move with first order velocity $\vc{U}_1(t)=\mathrm{Re}[\vc{U}_{1\omega}\exp(-i\omega t)]$. The first order flow velocity in the laboratory frame is related to that in the co-moving frame by
\begin{equation}
\label{2.10}\vc{u}^{(1)}(\vc{r},t)=\vc{v}^{(1)}(\vc{r},t)+\vc{U}_1(t).
\end{equation}
The displacement-distortion at the surface $r=a$ in the laboratory frame $\vc{\xi}'_\omega(\vc{s})$ is therefore given by
\begin{equation}
\label{2.11}-i\omega\vc{\xi}'_\omega(\vc{s})=-i\omega\vc{\xi}_\omega(\vc{s})+\vc{U}_{1\omega}.
\end{equation}
Here $\vc{s}$ denotes a labeled point on the surface of the undistorted sphere.
We show in the next section that for a uniform sphere the velocity $\vc{U}_{1\omega}$ can be calculated from the distortion $\vc{\xi}_\omega(\vc{s})$ by use of Newton's equation.

\section{\label{3}Dipolar distortion and first order motion}

The surface displacement-distortion function $\vc{\xi}'_\omega(\vc{s})$ can be expanded as
\begin{equation}
\label{3.1}\vc{\xi}'_\omega(\vc{s})=-ia\big[\kappa'_1\vc{v}_1(\vc{s},\alpha)+\mu'_1\vc{u}_1(\vc{s})+\sum^\infty_{l=2}[\kappa_l\vc{v}_l(\vc{s},\alpha)+\mu_l\vc{u}_l(\vc{s})]\big],
\end{equation}
with the same coefficients for $l\geq 2$ as in Eq. (2.9). We write the $l=1$ contribution to the velocity at the surface $r=a$ in the laboratory frame in terms of vector spherical harmonics as
\begin{equation}
\label{3.2}\vc{u}^{(1)}_{\omega 1}(\vc{s})=c_{A1}\vc{A}_1+c_{B1}\vc{B}_1.
\end{equation}
Here $\vc{A}_1=\vc{e}_z$ corresponds to displacement, and the vector spherical harmonic $\vc{B}_1=\vc{e}_z-3\cos\theta\;\vc{e}_r$ corresponds to a dipolar distortion. The corresponding $l=1$ contribution to the flow velocity takes the form
\begin{equation}
\label{3.3}\vc{u}^{(1)}_{\omega 1}(\vc{r})=\frac{1}{2}\alpha a c_{A1}\vc{v}_1(\vc{r},\alpha)+\bigg(\frac{k_2(\alpha a)}{2k_0(\alpha a)}\;c_{A1}-c_{B1}\bigg)\vc{u}_1(\vc{r}).
\end{equation}
The contribution to the first order pressure is
\begin{equation}
\label{3.4}p^{(1)}_{\omega 1}(\vc{r})=\eta\alpha^2 a\bigg(\frac{k_2(\alpha a)}{2k_0(\alpha a)}\;c_{A1}-c_{B1}\bigg)\bigg(\frac{a}{r}\bigg)^2\cos\theta.
\end{equation}
From the stress tensor we find for the first order force exerted by the fluid on the sphere $\vc{K}^{(1)}_{\omega}=K^{(1)}_{\omega}\vc{e}_z$ with
\begin{equation}
\label{3.5}K^{(1)}_{\omega}=\frac{1}{2}\alpha a c_{A1}k_{v1}(\alpha)+\bigg(\frac{k_2(\alpha a)}{2k_0(\alpha a)}\;c_{A1}-c_{B1}\bigg)k_{u1}(\alpha),
\end{equation}
with functions $k_{v1}(\alpha),k_{u1}(\alpha)$ given by
\begin{equation}
\label{3.6}k_{v1}(\alpha)=-8\pi\bigg(1+\frac{1}{\alpha a}\bigg)\eta a,\qquad k_{u1}(\alpha)=\frac{8\pi i}{3}s^2\eta a,
\end{equation}
with scale number $s=a\sqrt{\omega\rho/2\eta}$ and $\alpha=(s-is)/a$. The contribution $k_{v1}(\alpha)$ is purely viscous, and $k_{u1}(\alpha)$ arises from the pressure.

A solution of the above type occurred first in the derivation of the linear response of a rigid sphere immersed in a viscous incompressible fluid to an applied oscillatory force \cite{19},\cite{20}. In this solution
the flow velocity at the surface $r=a$ is $\vc{U}_{1\omega}=c_{A1}\vc{e}_z$, so that $c_{A1}=U_{1\omega},c_{B1}=0$. The equation of motion for the sphere with mass $m_0=4\pi\rho_0a^3/3$ reads in Fourier transform to first order in amplitude
\begin{equation}
\label{3.7}-i\omega m_0U_{1\omega}=K^{(1)}_{\omega}+E_\omega.
\end{equation}
where $K^{(1)}_{\omega}$ is the force exerted by the fluid on the sphere given by Eq. (3.5) and $E_\omega$ is the amplitude of the mechanical force applied to the sphere. Hence
one finds
\begin{equation}
\label{3.8}U_{1\omega}=\mathcal{Y}_t(\omega)E_\omega,
\end{equation}
where $\mathcal{Y}_t(\omega)$ is the translational admittance of a sphere with no-slip boundary condition, given by \cite{21}
\begin{equation}
\label{3.10}\mathcal{Y}_t(\omega)=\big[-i\omega(m_0+\frac{1}{2}m_f)+\zeta(\omega)\big]^{-1},\qquad m_f=\frac{4\pi}{3}\rho\;a^3,
\end{equation}
with frequency-dependent friction coefficient
\begin{equation}
\label{3.11}\zeta(\omega)=6\pi\eta a(1+\alpha a).
\end{equation}
Here the contribution $\frac{1}{2}m_f$ is the so-called added mass \cite{22}. In this solution the hydrodynamic force $K^{(1)}_{\omega}$ has both a viscous and a pressure contribution.

In a second type of solution we consider the first order flow generated by the displacement function in Eq. (2.9) with specified coefficients and $\kappa_1=0$. Here we use the equation of motion (3.7)
with $E_\omega=0$, so that the force exerted on the fluid is caused purely by oscillatory surface displacement.
By substitution of Eq. (3.5) into Eq. (3.7) we find a relation between the coefficients $c_{A1},c_{B1}$ of the form
\begin{equation}
\label{3.12}c_{A1}=\gamma_D(\omega)c_{B1},
\end{equation}
with the explicit expression
\begin{equation}
\label{3.13}\gamma_D(\omega)=-i\omega m_f\mathcal{Y}_t(\omega).
\end{equation}
We call $\gamma_D(\omega)$ the dipolar admittance, since it relates the convective motion of the sphere to the coefficient of dipolar distortion. The admittance may be expressed as
\begin{equation}
\label{3.14}\gamma_D(\omega)=\frac{4s^2\rho}{4s^2\rho_0+[9i+(9+9i)s+2s^2]\rho}.
\end{equation}

For known amplitude $c_{B1}$ of dipolar distortion the coefficient $c_{A1}$ follows from Eq. (3.12). This determines the amplitude of first order convective motion. In turn this determines the modification of the flow pattern from dipolar form, as given by Eq. (3.3). From a comparison of Eqs. (3.1) and (3.3) we find for the coefficients $\kappa'_1,\mu'_1$ by use of the no-slip boundary condition
\begin{equation}
\label{3.15}\kappa'_1=\frac{-\alpha}{2\omega}\; c_{A1},\qquad\mu'_1=\frac{-1}{\omega a}\bigg(\frac{k_2(\alpha a)}{2k_0(\alpha a)}\;c_{A1}-c_{B1}\bigg).
\end{equation}
Using the relation
\begin{equation}
\label{3.16}\vc{e}_z=\frac{1}{2}\alpha a\vc{v}_1(\vc{s},\alpha)+\frac{k_2(\alpha a)}{2k_0(\alpha a)}\vc{u}_1(\vc{s}),
\end{equation}
and comparing with Eq. (2.11) we find
\begin{equation}
\label{3.17}\kappa_1=0,\qquad\mu_1=\frac{1}{\omega a}\;c_{B1}.
\end{equation}
Given $\mu_1$ this yields $c_{B1}$. Hence we find the velocity $U_{1\omega}=c_{A1}$ by use of $c_{A1}=\gamma_D(\omega)c_{B1}$.

In this solution the hydrodynamic reaction force $K^{(1)}_{\omega}$ also has both a viscous and a pressure contribution. The relation between the two contributions follows from Eq.  (3.5) and the values of the two coefficients $c_{A1}$ and $c_{B1}$ for the given dipole moment $\mu_1$. The derivation shows that the flow pattern depends in an intricate way on viscosity, even though the primary dipolar flow is irrotational. In the Stokes limit $\omega=0$ the dipolar admittance vanishes, $\gamma_D(0)=0$, and then the first order velocity vanishes, $U_{10}=0$.

\section{\label{4}Mean swimming velocity and mean dissipation}

The coefficients $\kappa'_1,\mu'_1$ in Eq. (3.14) are used in the calculation of the mean swimming velocity, as discussed below.
In our earlier work on swimming of a sphere in a fluid with inertia \cite{2},\cite{15},\cite{16} we considered only strokes for which to first order in the amplitude of the stroke the sphere velocity vanishes. Recent study of the dipole-quadrupole (DQ) model \cite{1} made us realize that more general strokes must be considered. If the oscillatory flow pattern has a dipolar component, then the fluid exerts an oscillatory force on the sphere, and we must expect an oscillatory response velocity, which depends on the internal dynamics. Here we consider the simplest case, where the body is uniform. Then the response of the sphere depends on its mass density, as shown in Eq. (3.13), and consequently the first order flow pattern, as given by Eqs. (3.3) and (3.5), also does.

The first order velocity $\vc{U}_1(t)$ as calculated above affects also the first order flow pattern. The flow during the first period can be expressed in complex notation in terms of the amplitude functions as
\begin{eqnarray}
\label{4.1}\vc{u}^{(1)}_\omega(\vc{r})&=&-\omega a\bigg[\kappa'_1\vc{v}_1(\vc{r},\alpha)+\mu'_1\vc{u}_1(\vc{r})+\sum^\infty_{l=2}\big[\kappa_l\vc{v}_l(\vc{r},\alpha)+\mu_l\vc{u}_l(\vc{r})\big]\bigg],\nonumber\\
p^{(1)}_\omega(\vc{r})&=&-\eta\omega\alpha^2a^2\bigg[\mu'_1\bigg(\frac{a}{r}\bigg)^2\cos\theta+\sum^\infty_{l=2}\mu_l\bigg(\frac{a}{r}\bigg)^{l+1}P_l(\cos\theta)\bigg].
\end{eqnarray}
The mean second order flow velocity $\overline{\vc{v}^{(2)}}$ and pressure $\overline{p^{(2)}}$ in the rest frame, moving with mean swimming velocity $\overline{U^{(2)}}\vc{e}_z$ with respect to the laboratory frame, satisfy the inhomogeneous Stokes equations \cite{2}
\begin{equation}
\label{4.2}
\eta\nabla^2\overline{\vc{v}^{(2)}}-\nabla\overline{p^{(2)}}=\frac{1}{2}\rho\;\mathrm{Re}\;[\vc{u}^{(1)*}_\omega\cdot\nabla\vc{u}^{(1)}_\omega],\qquad\nabla\cdot\overline{\vc{v}^{(2)}}=0,
\end{equation}
with boundary condition
\begin{equation}
\label{4.3}
\overline{\vc{v}^{(2)}}\big|_{r=a}=\overline{\vc{u}}_S(\theta)=-\frac{1}{2}\mathrm{Re}\;[{\vc{\xi}'}^*_\omega\cdot\nabla\vc{u}^{(1)}_\omega]\bigg|_{r=a}.
\end{equation}
The mean is defined as the time-average over the period. The boundary condition is the time average of the no-slip condition calculated to second order in the amplitude of distortion,
\begin{equation}
\label{4.4}
\vc{v}^{(2)}\big|_{r=a}+\vc{\xi}'\cdot\nabla\vc{u}^{(1)}\bigg|_{r=a}=0,
\end{equation}
applied at the undistorted spherical surface. The right hand side in Eq. (4.2) represents the mean Reynolds force density $\overline{\vc{f}^{(2)}_R}=-\rho\overline{\vc{u}^{(1)}\cdot\nabla\vc{u}^{(1)}}$. We write $\overline{\vc{v}^{(2)}}=\overline{\vc{v}^{(2)}_{V}}+\overline{\vc{v}^{(2)}_{S}}$. The volume part of the second order flow $\overline{\vc{v}^{(2)}_{V}},\overline{p^{(2)}_{V}}$ satisfies Eq. (4.2) with the boundary condition $\overline{\vc{v}^{(2)}}_V\big|_{r=a}=0$. The surface part $\overline{\vc{v}^{(2)}_{S}},\overline{p^{(2)}_{S}}$ satisfies Eq. (4.2) with right hand side equal to zero and with boundary condition (4.3). The solution of Eq. (4.2) is determined by the condition that on time average no force is exerted on the fluid.

We define the multipole moment vectors $\vc{\psi}$ and $\hat{\vc{\psi}}$ as the one-dimensional arrays of complex coefficients
\begin{equation}
\label{4.5}\vc{\psi}=(\kappa'_1,\mu'_1,\kappa_2,\mu_2,....),\qquad\hat{\vc{\psi}}=(\mu_1,\kappa_2,\mu_2,....).
\end{equation}
At this point we regard the coefficients to be unknown.
The mean swimming velocity
$\overline{U^{(2)}}=\overline{U}_2$ and the mean rate of dissipation $\overline{\mathcal{D}^{(2)}}=\overline{\mathcal{D}}_2$ are bilinear in the coefficients. The bilinear dependence can be expressed with scalar products as
\begin{equation}
\label{4.6}\overline{U_2}=\frac{1}{2}\omega
a(\vc{\psi}|\du{B}|\vc{\psi}),\qquad\overline{\mathcal{D}_2}=8\pi\eta\omega^2a^3(\vc{\psi}|\du{A}|\vc{\psi}),
\end{equation}
with hermitian matrices $\du{B}$ and $\du{A}$. The matrix $\du{B}$ has non-vanishing elements only for indices corresponding to pairs of angular numbers $l,l-1$ and $l,l+1$, and the matrix $\du{A}$ has non-vanishing elements only for indices corresponding to the pair $l,l$.
In our earlier work \cite{16} we calculated
\begin{equation}
\label{4.7}\hat{\overline{U_2}}=\frac{1}{2}\omega
a(\hat{\vc{\psi}}|\hat{\du{B}}|\hat{\vc{\psi}}),\qquad\hat{\overline{\mathcal{D}_2}}=8\pi\eta\omega^2a^3(\hat{\vc{\psi}}|\hat{\du{A}}|\hat{\vc{\psi}}).
\end{equation}
with truncated matrices $\hat{\du{A}}$ and $\hat{\du{B}}$ obtained from $\du{A}$ and $\du{B}$  by dropping the first row and column. We can write
\begin{equation}
\label{4.8}\overline{U_2}=\hat{\overline{U_2}}+\overline{U_2}^{\;\prime},\qquad\overline{\mathcal{D}_2}=\hat{\overline{\mathcal{D}_2}}+\overline{\mathcal{D}_2}^{\;\prime},
\end{equation}
with correction terms $\overline{U_2}^{\;\prime}$ and $\overline{\mathcal{D}_2}^{\;\prime}$.

In Appendix B of Ref. 16 we provided explicit expressions for the matrix elements of $\hat{\du{A}}$ and $\hat{\du{B}}$ up to maximum angular number $l_{max}=L=3$. For our present purpose we need to evaluate in addition the elements $A_{00},A_{01},A_{10}$, where the subscript $0$ refers to the first element of a row or column, and the subscript $1$ refers to the second element, etc.. We also need elements $B_{02},B_{20},B_{03},B_{30}$.

The explicit expressions for the elements $A_{00},A_{01},A_{10}$ are
\begin{eqnarray}
\label{4.9}A_{00}&=&\frac{3}{8s^6}[9+18s+18s^2+12s^3+6s^4+2s^5],\nonumber\\
A_{01}&=&A_{10}^*=\frac{3}{4s^3}[3+3i+6is-(2-2i)s^2].
\end{eqnarray}
The non-vanishing values of the other elements are listed in Appendix B of Ref. 16.

The matrix $\du{B}$ is conveniently written as a sum of two parts
\begin{equation}
\label{4.10}\du{B}=\du{B}_S+\du{B}_B,
\end{equation}
where $\du{B}_S$ follows directly from the second order surface velocity $\overline{\vc{u}}_S(\theta)$, and $\du{B}_B$ follows from the Reynolds force density. The calculations are performed in the same manner as before \cite{16}.
The explicit expressions for the elements $B_{S02},B_{S20},B_{S03},B_{S30}$ are
\begin{eqnarray}
\label{4.11}B_{S02}&=&B_{S20}^*=\frac{3-3i}{8s^7}\big[45+90s+(90-6i)s^2+(60-12i)s^3+(28-12i)s^4+(8-8i)s^5\big],\nonumber\\
B_{S03}&=&B_{S30}^*=\frac{-3i}{20s^3}\big[15+15i+30is-12(1-i)s^2-4s^3\big].
\end{eqnarray}
The explicit expressions for the elements $B_{B02},B_{B20},B_{B03},B_{B30}$ are
\begin{eqnarray}
\label{4.12}B_{B02}=B_{B20}^*&=&\frac{-1-i}{16s^3}\big[9i+18is+(6+18i)s^2-(4+36i)s^3\nonumber\\&+&4s^4-8s^5+8s^4(9i+2s^2)e^{2s}\Gamma(0,2s)\big],\nonumber\\
B_{B03}=B_{B30}^*&=&\frac{-s-is}{120}\big[3i-(3-3i) s+6s^2-(8+8i)s^3\nonumber\\
&-&is^4-(1-i)s^5+2s^4(9i+s^2)e^{s+is}\Gamma(0,s+is)\big],
\end{eqnarray}
where $\Gamma(0,z)$ is an incomplete Gamma-function \cite{17}. All other elements of the matrices vanish.

\section{\label{5}Stokes representation}

The matrices $\du{A}(s)$ and $\du{B}(s)$ are singular at $s=0$ which causes difficulties in numerical calculations and in the discussion of the relation to swimming in the Stokes limit. As we showed earlier \cite{3}, we can choose a more convenient matrix representation by expanding the surface displacement-distortion $\vc{\xi}'_\omega(\vc{s})$ in terms of a different set of vector functions defined on the surface of the sphere $r=a$. It is of particular interest to use the set of functions found as limiting values on the sphere surface of the modes defined in the Stokes limit \cite{3}. The mode functions $\vc{u}_l(\vc{r})$ in the Stokes limit are the same as in Eq. (2.5), but the functions $\vc{v}_l(\vc{r},\alpha)$ are changed to
\begin{eqnarray}
\label{5.1}\vc{v}^0_l(\vc{r})&=&\bigg(\frac{a}{r}\bigg)^l\bigg[(l+1)P_l(\cos\theta)\vc{e}_r+\frac{l-2}{l}P^1_l(\cos\theta)\vc{e}_\theta\bigg]\nonumber\\
&=&\bigg(\frac{a}{r}\bigg)^l\bigg[\frac{2l+2}{l(2l+1)}\vc{A}_l-\frac{2l-1}{2l+1}\vc{B}_l\bigg].
\end{eqnarray}
We denote the corresponding set of superposition coefficients as $\vc{\psi}^I=(\kappa_1'^I,\mu_1'^I,\kappa_2^I,\mu_2^I,...)$ and the corresponding Stokes representation of the matrices as $\du{A}^I(s)$ and $\du{B}^I(s)$. The Stokes limit is denoted as $\du{A}^0=\du{A}^I(0)$ and $\du{B}^0=\du{B}^I(0)$.

The two sets of mode coefficients  $\vc{\psi}$ and $\vc{\psi}^I$ in the two representations are related by
\begin{equation}
\label{5.2}\vc{\psi}=\du{T}\cdot\vc{\psi}^I,
\end{equation}
with a transformation matrix $\du{T}$. The matrix $\du{T}$ is block-diagonal, as given by a factor $\delta_{ll'}$, with a 2-dimensional $\du{T}_l$ at order $l$ given by the relations
\begin{eqnarray}
\label{5.3}\kappa_l&=&\frac{\pi}{l(2l+1)e^zk_{l-1}(z)}\;\kappa^I_l,\nonumber\\
\mu_l&=&\frac{1}{2l+1}\bigg[2l-1+2\frac{k_{l+1}(z)}{k_{l-1}(z)}\bigg]\kappa^I_l+\mu^I_l,\qquad z=(1-i)s.
\end{eqnarray}
The relation between the two sets of matrices is
\begin{equation}
\label{5.4}\du{A}^I=\du{T}^\dagger\cdot\du{A}\cdot\du{T},\qquad\du{B}^I=\du{T}^\dagger\cdot\du{B}\cdot\du{T},
\end{equation}
where $\du{T}^\dagger$ is the hermitian conjugate of $\du{T}$. The mean swimming velocity
$\overline{U}_2$ and the mean rate of dissipation $\overline{\mathcal{D}}_2$ can be expressed alternatively as
\begin{equation}
\label{5.5}\overline{U_2}=\frac{1}{2}\omega
a(\vc{\psi}^I|\du{B}^I|\vc{\psi}^I),\qquad\overline{\mathcal{D}_2}=8\pi\eta\omega^2a^3
(\vc{\psi}^I|\du{A}^I|\vc{\psi}^I).
\end{equation}

Earlier \cite{3} we gave the expressions of the matrix elements of the truncated matrices $\hat{\du{A}}^I$ and $\hat{\du{B}}^I$ up to order $l=3$. The missing elements of the matrix $\du{A}^I$ read
\begin{equation}
\label{5.6}A^I_{00}=1+\frac{2}{3}s,\qquad A^I_{01}=A^I_{10}=1.
\end{equation}
All other elements of the first row and column vanish. The matrix $\du{B}^I$ is a sum of two terms $\du{B}^I=\du{B}^I_S+\du{B}^I_B$. The missing elements of the matrix $\du{B}^I_S$ read
 \begin{eqnarray}
\label{5.7}B^I_{S02}&=&B^{I*}_{S20}=\frac{-1}{5}\frac{1+(3+i)s+(4-4i)s^2}{i+s+is},\nonumber\\
B^I_{S03}&=&B^{I*}_{S30}=\frac{-i}{5}[3-(2+2i)s].
\end{eqnarray}

The missing elements $B^I_{B02}$ and $B^I_{B20}$ are given by
 \begin{eqnarray}
\label{5.8}B^I_{B02}&=&B^{I*}_{B20}=\frac{-is^2}{90(1+s-is)}\bigg[-27i-54s+(39-18i)s^2-(36-6i)s^3-(22+3i)s^4\nonumber\\
&-&(4-2i)s^5+2is^6-
s^2\bigg(54+(54-54i)s-60is^2-(24+24i)s^3-6s^4-2(1-i)s^5\bigg)F_+\nonumber\\
&-&36s^2[1+s+is+is^2]F_-+24s^2(9-2is^2)F_2\bigg],
\end{eqnarray}
with the abbreviations
\begin{equation}
\label{5.9}F_+=F(s+is),\qquad F_-=F(s-is),\qquad F_2=F(2s),
\end{equation}
where the function $F(z)$ with complex variable $z$ is defined by
\begin{equation}
\label{5.10}F(z)=e^zE_1(z)=\int^\infty_0\frac{e^{-u}}{z+u}\;du.
\end{equation}
The missing elements $B^I_{B03}$ and $B^I_{B30}$ are given by
 \begin{eqnarray}
\label{5.11}B^I_{B03}=B^{I*}_{B30}=\frac{s^2}{90}&\big[&3+(3+3i)s-6is^2-(8-8i)s^3-s^4\nonumber\\&+&(1+i)s^5+2s^4(9-is^2)F_+\big].
\end{eqnarray}
The elements are found from the matrices listed in Sec. 4 by use of the transformation (5.4). Elements which are not listed vanish.

\section{\label{6}Mean swimming velocity and mean rate of dissipation for some simple strokes}

The first order center of mass motion has an effect on the mean swimming velocity and the mean rate of dissipation given by the correction terms in Eq. (4.8). For a sphere only the dipolar surface distortion leads to oscillatory motion. Higher order multipole moments of the distortion do not couple by symmetry. In this section we consider the effect of first order motion on the mean swimming velocity and on the mean dissipation for five simple swimmers studied before in Ref. 3. In addition we consider also the optimal swimming stroke for multipoles of order $l=1,2,3$.

As before we define the dimensionless reduced mean swimming velocity as \cite{3}
\begin{equation}
\label{6.1}U_{red}(s)=\frac{(\vc{\psi}^I|\du{B}^I(s)|\vc{\psi}^I)}{(\hat{\vc{\psi}}^I|\hat{\du{A}}^0|\hat{\vc{\psi}}^I)}.
\end{equation}
The denominator provides a measure of the intensity of surface agitation. Here the vector $\hat{\vc{\psi}}^I$ is given by
\begin{equation}
\label{6.2}\hat{\vc{\psi}}^I=(\mu^I_1,\kappa^I_2,\mu^I_2,....),
\end{equation}
with a chosen set of coefficients. The vector $\hat{\vc{\psi}}^I$ determines the surface deformation $\vc{\xi}(\theta,t)$ as a harmonic function of time. We consider five different sets of coefficients $\hat{\vc{\psi}}^I$ independent of $s$, corresponding to five different strokes in fluids characterized by scaling parameter $s$.

 The vector
\begin{equation}
\label{6.3}\vc{\psi}^I=(\kappa^I_1(s),\mu^I_1,\kappa^I_2,\mu^I_2,....)
\end{equation}
is determined from Eqs. (3.14) and (5.3) by use of $c_{A1}=\gamma_D(\omega)c_{B1}$. Here Eq. (5.3) is used for $l=1$ with left-hand side given by $\kappa'_1,\mu'_1$. This yields
\begin{equation}
\label{6.4}\kappa^I_1(s)=\frac{-3s^2\rho}{4s^2\rho_0+[9i+(9+9i)s+3s^2]\rho}\;\mu^I_1.
\end{equation}
The moment $\kappa^I_1(s)$ may be regarded as being induced by $\mu^I_1$. The moment gives the amplitude of first order linear motion corresponding to the dipole moment $\mu^I_1$ by the mechanism discussed in Sec. 3. We compare $U_{red}(s)$ with the previously defined quantity \cite{3}
\begin{equation}
\label{6.5}\hat{U}_{red}(s)=\frac{(\hat{\vc{\psi}}^I|\hat{\du{B}}^I(s)|\hat{\vc{\psi}}^I)}{(\hat{\vc{\psi}}^I|\hat{\du{A}}^0|\hat{\vc{\psi}}^I)},
\end{equation}
which has a numerator different from the one in Eq. (6.1). Correspondingly we define the reduced mean rate of dissipation as
\begin{equation}
\label{6.6}\mathcal{D}_{red}(s)=\frac{(\vc{\psi}^I|\du{A}^I(s)|\vc{\psi}^I)}{(\hat{\vc{\psi}}^I|\hat{\du{A}}^0|\hat{\vc{\psi}}^I)}.
\end{equation}
This can be compared with $\hat{\mathcal{D}}_{red}(s)$, defined in analogy to Eq. (6.5). It is also of interest to consider the efficiency $L_T(s)$, defined by
\begin{equation}
\label{6.7}L_T(s)=2\pi E_T(s)=8\pi\eta\omega a^2\frac{|U(s)|}{\mathcal{D}(s)}=\frac{1}{2}\frac{|U_{red}(s)|}{\mathcal{D}_{red}(s)}.
\end{equation}
Originally \cite{2} we defined $E_T(s)$, but showed later \cite{23} that it is advantageous to introduce the numerical factor $2\pi$.
The coefficient $L_T(s)$ can be compared with $\hat{L}_T(s)$ defined in the same way from $|\hat{U}_{red}(s)|$ and $\hat{\mathcal{D}}_{red}(s)$.

We consider first the Stokes limit $s=0$. The upper left-hand corner of the matrices $\du{A}^0=\du{A}^I(0)$ and $\du{B}^0=\du{B}^I(0)$, truncated at $l=4$, is given by equations (7.17) and (7.11) of Ref. 23. We have $U_{red}(0)=\hat{U}_{red}(0)$, since $\kappa^I_1(0)=0$. The value $U_{red}(0)$ can be evaluated for chosen stroke from the explicit matrix form. \\
Next we consider the inertia limit $s\rightarrow\infty$. The upper left-hand corner of the matrix $\du{B}^I(\infty)$, truncated at $l=3$ is given by
\begin{equation}
\label{6.8}\du{B}^I_{13}(\infty)=\left(\begin{array}{cccccc}0&0&\frac{2+23i}{5}&\frac{17i}{5}&0&0
\\0&0&\frac{3i}{5}&-3i&0&0
\\\frac{2-23i}{5}&\frac{-3i}{5}&0&0&\frac{8+110i}{35}&\frac{6i}{7}
\\\frac{-17i}{5}&3i&0&0&\frac{-58i}{35}&-6i
\\0&0&\frac{8-110i}{35}&\frac{58i}{35}&0&0
\\0&0&\frac{-6i}{7}&6i&0&0
\end{array}\right).
\end{equation}
The subscripts $13$ indicate that angular numbers $l=1,2,3$ are involved. The matrix $\du{B}^I_{13}(\infty)$ may be used to calculate the mean swimming velocity in the limit of large $s$ for general stroke $\hat{\vc{\psi}}^I=(\mu^I_1,\kappa^I_2,\mu^I_2,\kappa^I_3,\mu^I_3)$, together with
\begin{equation}
\label{6.9}\kappa^I_1(\infty)=\frac{-3\rho}{4\rho_0+3\rho}\;\mu^I_1.
\end{equation}
The corresponding surface agitation $(\hat{\vc{\psi}}^I|\hat{\du{A}}^0|\hat{\vc{\psi}}^I)$ follows from the matrix $\hat{\du{A}}^0_{13}$. From these expressions we can evaluate the limiting values $U_{red}(\infty)$ and $\hat{U}_{red}(\infty)$ in the examples given below.

In our first example we consider the swimmer with a dipolar and a quadrupolar flow field, corresponding to moments $\mu^I_1=1$, $\mu^I_2=i/\sqrt{2}$, and all other moments vanishing. The dipolar and quadrupolar flow fields vary harmonically in time, and out of phase. The surface shape at sixteen equidistant instants in a period is shown in Fig. 1. For this swimmer the primary reduced swimming velocity $\hat{U}_{red}=1/\sqrt{2}=0.701$, independent of $s$. In Fig. 2 we compare $U_{red}(s)$ with $\hat{U}_{red}$ as a function of $s$ for $\rho_0=\rho$. In the Stokes limit $s\rightarrow 0$ the two quantities become identical. The comparison shows the effect of the linear motion on the mean swimming velocity. In situations without the linear motion the mean swimming velocity is given by $\hat{U}_{red}$. For a uniform sphere the reaction force leads to $U_{red}(s)$. For a sphere with different internal dynamics the behavior will be intermediate to the two curves shown in Fig. 2.

For this stroke the limiting value of $U_{red}(s)$ at large $s$ for general $\rho_0$ is given by
\begin{equation}
\label{6.10}U_{red}(\infty)=\frac{2\sqrt{2}}{5}\;\frac{5\rho_0+8\rho}{4\rho_0+3\rho}.
\end{equation}
For $\rho_0=\rho$ this takes the value $26\sqrt{2}/35=1.051$.

In Fig. 3 we show the efficiency $L_T(s)$ of the stroke. For this stroke the efficiency of the swimmer without reaction force is $\hat{L}_T=0.354$, independent of $s$. At $s=10$ the efficiency of the swimmer with reaction force is $L_T(10)=0.409$.

In Fig. 4 we compare $U_{red}(s)$ with $\hat{U}_{red}(s)$ as a function of $s$ for $\rho_0=\rho$ for the swimmer with stroke characterized by
\begin{equation}
\label{6.11}\mu^I_1=1,\qquad\kappa^I_2=-\mu^I_2,\qquad\mu^I_2=\frac{5}{3}i.
\end{equation}
The surface shape at sixteen equidistant instants in a period is shown in Fig. 5. This swimmer is a so-called $B_1B_2$-squirmer with $B_2/B_1=5$. Ishikawa et al. \cite{24} considered an active particle characterized by these coefficients. In the Stokes limit $U_{red}(s)$ and $\hat{U}_{red}(s)$ both tend to $48/43$. For this swimmer the limiting value of $U_{red}(s)$ at large $s$ for general $\rho_0$ is given by
\begin{equation}
\label{6.12}U_{red}(\infty)=\frac{144}{43}\;\frac{2\rho_0+\rho}{4\rho_0+3\rho}.
\end{equation}
For $\rho_0=\rho$ this takes the value $432/301=1.435$. For this stroke the efficiency of the swimmer without reaction force is $\hat{L}_T(10)=0.371$ at $s=10$. The efficiency of the swimmer with reaction force is $L_T(10)=0.306$.

In Fig. 6 we compare $U_{red}(s)$ with $\hat{U}_{red}(s)$ as a function of $s$ for $\rho_0=\rho$ for the swimmer with stroke characterized by
\begin{equation}
\label{6.13}\mu^I_1=1,\qquad\kappa^I_2=-\frac{4i\sqrt{2}}{3},\qquad\mu^I_2=\frac{11i}{5\sqrt{2}}.
\end{equation}
The surface shape at sixteen equidistant instants in a period is shown in Fig. 7. In the Stokes limit $U_{red}$ and $\hat{U}_{red}$ both tend to $5/(3\sqrt{2})=1.179$. For this swimmer the limiting value of $U_{red}(s)$ at large $s$ for general $\rho_0$ is given by
\begin{equation}
\label{6.14}U_{red}(\infty)=\frac{2\sqrt{2}}{75}\;\frac{205\rho_0+64\rho}{4\rho_0+3\rho}.
\end{equation}
For $\rho_0=\rho$ this takes the value $538\sqrt{2}/525=1.449$.  For this stroke the efficiency of the swimmer without reaction force is $\hat{L}_T(10)=0.331$ at $s=10$. The efficiency of the swimmer with reaction force is $L_T(10)=0.250$.

In Fig. 8 we show the behavior the mean rate of dissipation, given by Eq. (6.6), for $\rho_0=\rho$. In each case the mean rate increases in proportion to $s$ for large $s$. The efficiency tends to zero in proportion to $1/s$.

In Fig. 9 we compare $U_{red}(s)$ with $\hat{U}_{red}(s)$ as a function of $s$ for $\rho_0=\rho$ for the swimmer characterized by \cite{25}
\begin{equation}
\label{6.15}\mu^I_1=1,\qquad\kappa^I_2=\frac{5}{3}\sqrt{\frac{230}{413}}\;i,\qquad\mu^I_2=0,\qquad
\kappa^I_3=-\frac{27}{59},\qquad\mu^I_3=0.
\end{equation}
The surface shape at sixteen equidistant instants in a period is shown in Fig. 10. Both $U_{red}(s)$ and $\hat{U}_{red}(s)$ change sign at some value of $s$. A sign change of the mean swimming velocity as a function of scale number $s$ was also observed in computer simulations of a two-sphere model by Jones et al. \cite{4}, Dombrowski et al. \cite{5}, and Dombrowski and Klotsa \cite{6}. These authors also found a center of mass velocity oscillating at frequency $\omega$. We presume that the mechanism is similar to that of the model considered here. For this stroke the efficiency of the swimmer without reaction force is $\hat{L}_T(10)=0.103$ at $s=10$. The efficiency of the swimmer with reaction force is $L_T(10)=0.011$.

In Fig. 11 we compare $U_{red}(s)$ with $\hat{U}_{red}(s)$ as a function of $s$  for $\rho_0=\rho$ for the swimmer characterized by
\begin{equation}
\label{6.16}\mu^I_1=1,\qquad\kappa^I_2=-1.553i,\qquad\mu^I_2=1.824i,\qquad
\kappa^I_3=1.373,\qquad\mu^I_3=-1.440.
\end{equation}
The surface shape at sixteen equidistant instants in a period is shown in Fig. 12. For this swimmer $U_{red}(s)$ and $\hat{U}_{red}(s)$ are both positive. In the limit $U_{red}(\infty)=2.109$ for $\rho_0=\rho$. For this stroke the efficiency of the swimmer without reaction force is $\hat{L}_T(10)=0.535$ at $s=10$. The efficiency of the swimmer with reaction force is $L_T(10)=0.499$.

Finally we consider the effect of the reaction force for the optimal stroke with multipoles of orders $l=1,2,3$. If the reaction force is completely absorbed with zero effect on the motion, then the optimal stroke follows from a generalized eigenvalue problem \cite{2}. In the Stokes representation this reads
\begin{equation}
\label{6.17}\hat{\du{B}}^I\hat{\psi}^I_\lambda=\lambda\hat{\du{A}}^I\hat{\psi}^I_\lambda,
\end{equation}
where $\hat{\du{B}}^I$ is the $5\times 5$ matrix obtained from $\du{B}^I(s)$ by deleting the first row and column, and similarly for $\hat{\du{A}}^I$. The matrix elements depend only on the scale number $s$. There are two pairs of eigenvectors of the form $(1,p,q,u,v)$ and $(1,-p,-q,u,v)$ with eigenvalues $\pm\lambda_+$. The eigenvectors of a conjugate pair correspond to swimming in opposite directions. The fifth eigenvector has the form $(1,0,0,u,v)$ and corresponds to eigenvalue zero.
In Fig. 13 we show the stroke corresponding to the maximum positive eigenvalue for $s=10$. The body swims to the right. The efficiency is optimal and equals $\hat{L}_{Tmax}(10)=0.697$.

We compare this with swimming with account of reaction force as in Sec. 3. The first two components of the stroke vector $\psi^I$ can be taken to be $\kappa^I_1(s)$ as in Eq. (6.4) and $\mu^I_1=1$, so that $\psi^I=(\kappa^I_1(s),1,p,q,u,v)$. We can find the vector of maximum efficiency $L_T$ by successive variation of the components $(p,q,u,v)$, looking for the optimum efficiency at each step. The iterative procedure converges fairly rapidly. 

In Fig. 14 we compare the optimal efficiencies $\hat{L}_{Tmax}$ and $L_{Tmax}$ as functions of scale number $s$ for $\rho_0=\rho$. For small $s$ the factor in $\kappa^I_1(s)$, as given by Eq. (6.4), tends to zero, so that in this limit the constraint has no effect and the two curves tend to coincide. For $s=1$ we find $\hat{L}_{Tmaz}(1)=0.758$ and $L_{Tmax}(1)=0.755$. In Fig. 15 we show the stroke for $s=10,\;\rho_0=\rho,$ and swimming to the right with reaction force acting. The corresponding efficiency is $L_{Tmax}(10)=0.693$. This is to be compared to $\hat{L}_{Tmax}(10)=0.697$ for swimming without reaction force. Note that these values are quite a bit higher than those in the examples above. With increasing $s$ the difference between the two types of swimming increases. At $s=100$ and $\rho_0=\rho$ we find $L_{Tmax}(100)=0.483$, compared with $\hat{L}_{Tmax}(100)=0.564$.

From the figures shown it is evident that it is difficult to characterize the effect of fluid inertia on the mean swimming velocity and the mean dissipation qualitatively. It is also hard to understand the effect of the reaction force for given stroke qualitatively. In both cases the explicit calculation is required to find the effect.

\section{\label{7}Discussion}

In the theory of swimming developed earlier \cite{2} the mean swimming velocity was calculated from the condition that on time average over a period the body with periodically changing shape does not exert  a force on the fluid.  The resulting mean swimming velocity is of second order in the amplitude of the stroke. In the derivation we assumed that to first order the body does not move. However, later we found by explicit calculation that for a sphere with dipolar surface distortion the sphere does exert a periodic force on the fluid, to first order in the amplitude of distortion and with zero mean \cite{1}. Conversely, this implies that in this case the sphere gets accelerated by a time-periodic reaction force. In the present calculation we assumed that the reaction force gets fully absorbed by center of mass motion, as is the case for a sphere with uniform mass density.

The key expression at the basis of the present calculation is given by Eq. (3.5). This shows that both a simple periodic translation and a dipolar distortion give rise to a reaction force acting on the sphere. The first case was already considered by Stokes many years ago \cite{19}, the second was studied only recently \cite{1}. For a uniform deformable sphere the first order effect on the motion of the sphere is determined from Newton's equation.

The conclusion is more general. If the inner dynamics of the swimmer is such that the first order reaction force has an effect on the body motion, then this must be taken into account, and its second order effect on the mean swimming velocity must be calculated. A purely kinematic theory, as we developed earlier \cite{2}, is valid only if the first order reaction force is fully absorbed and has no effect on the body motion.

In situations where the first order reaction force does have an effect on the body motion, then in a complete theory the inner dynamics of the body must be considered. For a uniform body this can be circumvented by the requirement that the net surface motion is identical with that of the center of mass. For a uniform deformable sphere this leads to a calculation of the dipolar admittance, as shown in Sec. 3. The subsequent calculation in Secs. 4-6 demonstrates that fluid inertia has an effect on the mean swimming velocity, as was suggested earlier \cite{2},\cite{26},\cite{27}.

Our theory of swimming is based on a perturbation theory to second order in the amplitude of the stroke. It is assumed that the flow remains laminar and that turbulence can be neglected in the full range of scale number. The assumption is well supported by the recent computer simulations of a two-sphere system by Jones et al. \cite{4}, Dombrowski et al. \cite{5}, and Dombrowski and Klotsa \cite{6}. In actual swimming the mean swimming velocity will be somewhat reduced by turbulent drag. For further confirmation of the theory it would be desirable to carry out computer simulations for a deformable sphere \cite{28}.

\newpage

\newpage

\section*{Figure captions}

\subsection*{Figure 1}
Plot of the surface shape at sixteen equidistant instants in a period for the dipole-quadrupole swimmer with stroke characterized by mode coefficients $\mu^I_1=1,\mu^I_2=i/\sqrt{2}$.
The figure is to be read from left to right and from top to bottom.

\subsection*{Figure 2}
Plot of the reduced mean swimming velocity $U_{red}(s)$ as a function of scale number $s$ for the dipole-quadrupole swimmer with $\rho_0=\rho$ and stroke characterized by mode coefficients $\mu^I_1=1,\mu^I_2=i/\sqrt{2}$ (solid curve) compared with $\hat{U}_{red}(s)$ for the same stroke of a swimmer without linear motion (dashed curve).

\subsection*{Figure 3}
Plot of the efficiency $L_T(s)$ for the dipole-quadrupole swimmer with account of reaction force (solid curve), compared with $\hat{L}_T(s)$ for fully absorbed reaction force (dashed curve).

\subsection*{Figure 4}
Plot of the reduced mean swimming velocity $U_{red}(s)$ as a function of scale number $s$ for the swimmer with $\rho_0=\rho$ and stroke characterized by mode coefficients specified in Eq. (6.11) (solid curve) compared with $\hat{U}_{red}(s)$ for the same stroke of a swimmer without linear motion (dashed curve).

\subsection*{Figure 5}
Plot of the surface shape at sixteen equidistant instants in a period for the swimmer with stroke characterized by mode coefficients specified in Eq. (6.11).

\subsection*{Figure 6}
Plot of the reduced mean swimming velocity $U_{red}(s)$ as a function of scale number $s$ for the swimmer with $\rho_0=\rho$ and stroke characterized by mode coefficients specified in Eq. (6.13) (solid curve) compared with $\hat{U}_{red}(s)$ for the same stroke of a swimmer without linear motion (dashed curve).

\subsection*{Figure 7}
Plot of the surface shape at sixteen equidistant instants in a period for the swimmer with stroke characterized by mode coefficients specified in Eq. (6.13).

\subsection*{Figure 8}
Plot of the reduced mean rate of dissipation $\mathcal{D}_{red}(s)$ as a function of scale number $s$ for the swimmer with $\rho_0=\rho$ for the dipole-quadrupole swimmer of Fig. 2 (solid curve), for the swimmer with stroke (6.11) (long dashes), and for the swimmer with stroke (6.13) (short dashes).

\subsection*{Figure 9}
Plot of the reduced mean swimming velocity $U_{red}(s)$ as a function of scale number $s$ for the swimmer with $\rho_0=\rho$ and stroke characterized by mode coefficients specified in Eq. (6.15) (solid curve) compared with $\hat{U}_{red}(s)$ for the same stroke of a swimmer without linear motion (dashed curve).

\subsection*{Figure 10}
Plot of the surface shape at sixteen equidistant instants in a period for the swimmer with stroke characterized by mode coefficients specified in Eq. (6.15).

\subsection*{Figure 11}
Plot of the reduced mean swimming velocity $U_{red}(s)$ as a function of scale number $s$ for the swimmer with $\rho_0=\rho$ and stroke characterized by mode coefficients specified in Eq. (6.16) (solid curve) compared with $\hat{U}_{red}(s)$ for the same stroke of a swimmer without linear motion (dashed curve).

\subsection*{Figure 12}
Plot of the surface shape at sixteen equidistant instants in a period for the swimmer with stroke characterized by mode coefficients specified in Eq. (6.16).

\subsection*{Figure 13}
Plot of the surface shape at sixteen equidistant instants in a period for the optimal swimmer with fully absorbed reaction force at $s=10$.

\subsection*{Figure 14}
Plot of the efficiency $\hat{L}_{Tmax}(s)$ of the optimal swimmer with fully absorbed reaction force (dashed curve), and of the efficiency $L_{Tmax}(s)$ of the optimal swimmer with account of reaction force via Newton's equation of motion for $\rho_0=\rho$ (solid curve), as functions of scale number $s$.

\subsection*{Figure 15}
Plot of the surface shape at sixteen equidistant instants in a period for the optimal swimmer with account of reaction force at $s=10,\;\rho_0=\rho$.

\newpage
\setlength{\unitlength}{1cm}
\begin{figure}
 \includegraphics{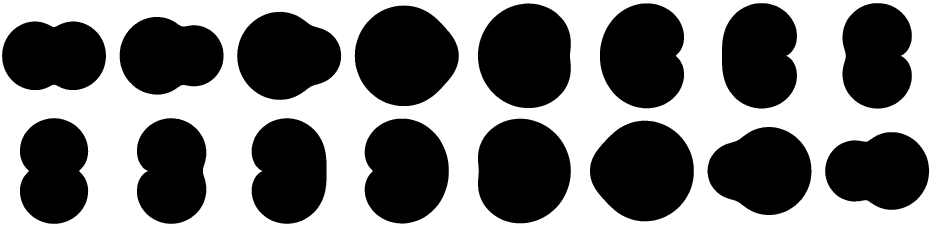}
   \put(-9.1,3.1){}
\put(-1.2,-.2){}
  \caption{}
\end{figure}
\newpage
\clearpage
\newpage
\setlength{\unitlength}{1cm}
\begin{figure}
 \includegraphics{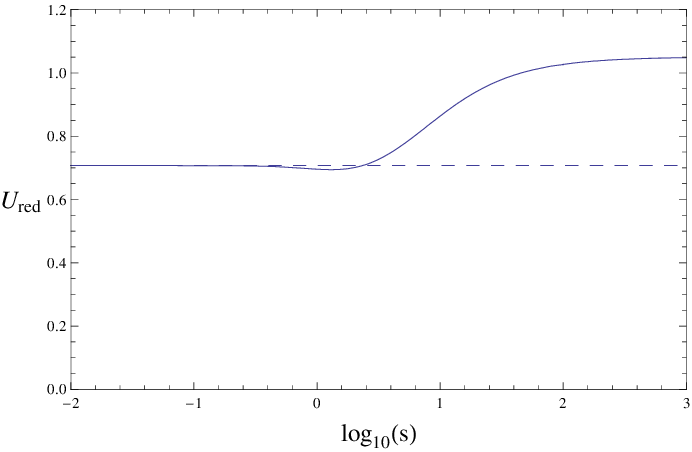}
   \put(-9.1,3.1){}
\put(-1.2,-.2){}
  \caption{}
\end{figure}
\newpage
\clearpage
\newpage
\setlength{\unitlength}{1cm}
\begin{figure}
 \includegraphics{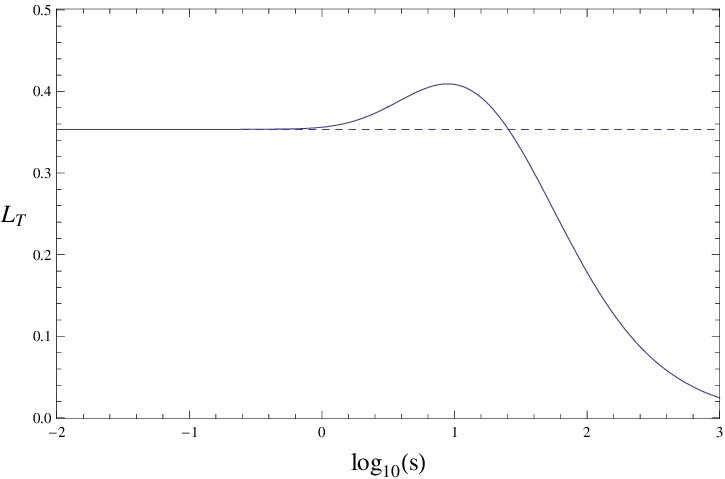}
   \put(-9.1,3.1){}
\put(-1.2,-.2){}
  \caption{}
\end{figure}
\newpage
\clearpage
\newpage
\setlength{\unitlength}{1cm}
\begin{figure}
 \includegraphics{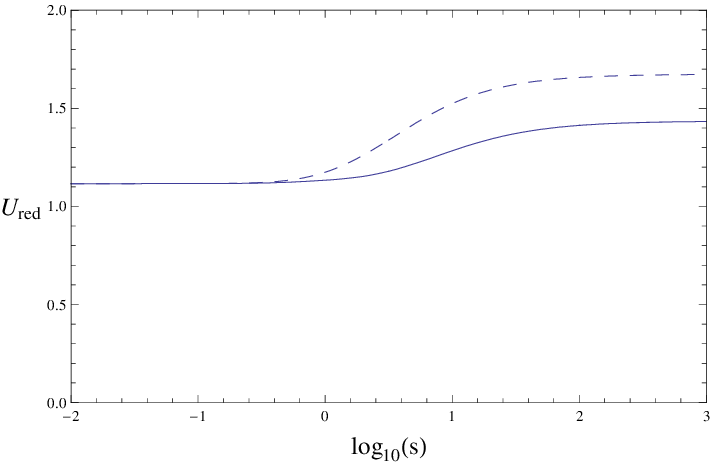}
   \put(-9.1,3.1){}
\put(-1.2,-.2){}
  \caption{}
\end{figure}
\newpage
\clearpage
\newpage
\setlength{\unitlength}{1cm}
\begin{figure}
 \includegraphics{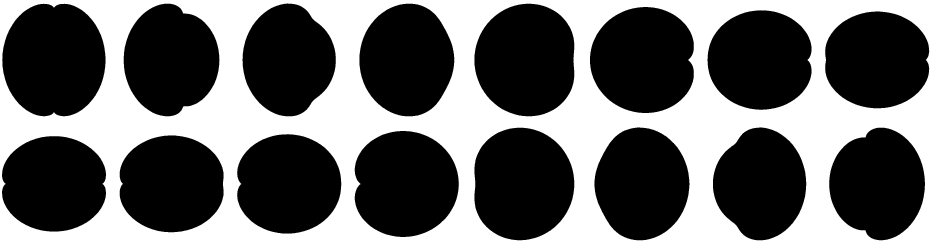}
   \put(-9.1,3.1){}
\put(-1.2,-.2){}
  \caption{}
\end{figure}
\newpage
\clearpage
\newpage
\setlength{\unitlength}{1cm}
\begin{figure}
 \includegraphics{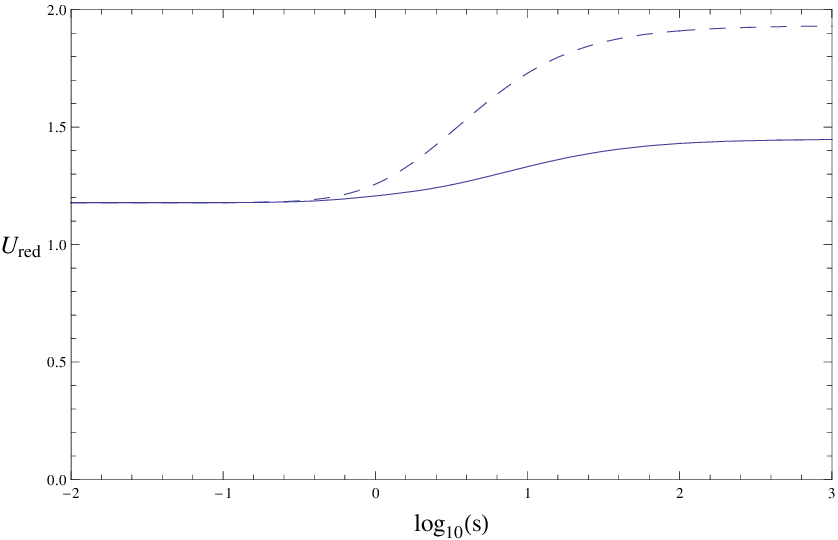}
   \put(-9.1,3.1){}
\put(-1.2,-.2){}
  \caption{}
\end{figure}
\newpage
\clearpage
\newpage
\setlength{\unitlength}{1cm}
\begin{figure}
 \includegraphics{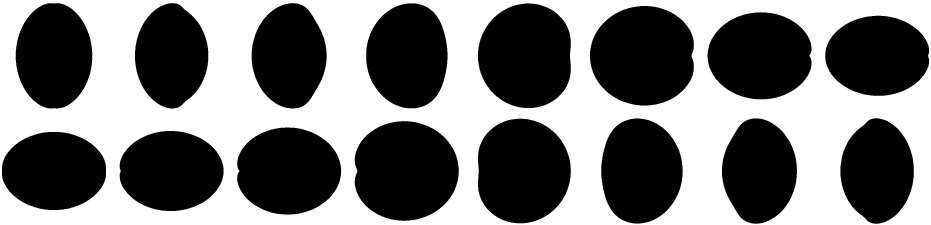}
   \put(-9.1,3.1){}
\put(-1.2,-.2){}
  \caption{}
\end{figure}
\newpage
\clearpage
\newpage
\setlength{\unitlength}{1cm}
\begin{figure}
 \includegraphics{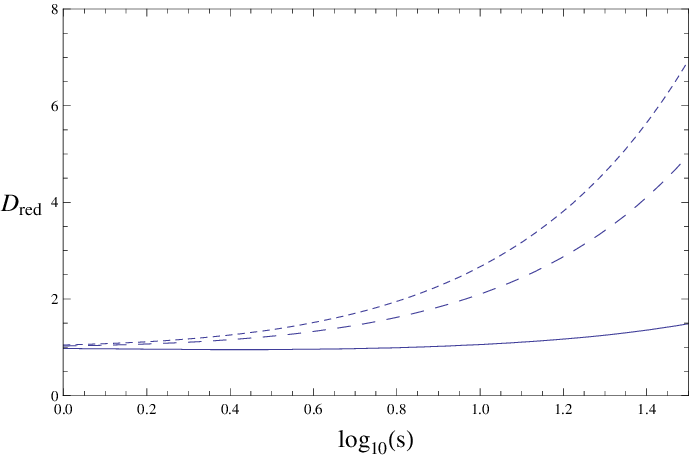}
   \put(-9.1,3.1){}
\put(-1.2,-.2){}
  \caption{}
\end{figure}
\newpage
\clearpage
\newpage
\setlength{\unitlength}{1cm}
\begin{figure}
 \includegraphics{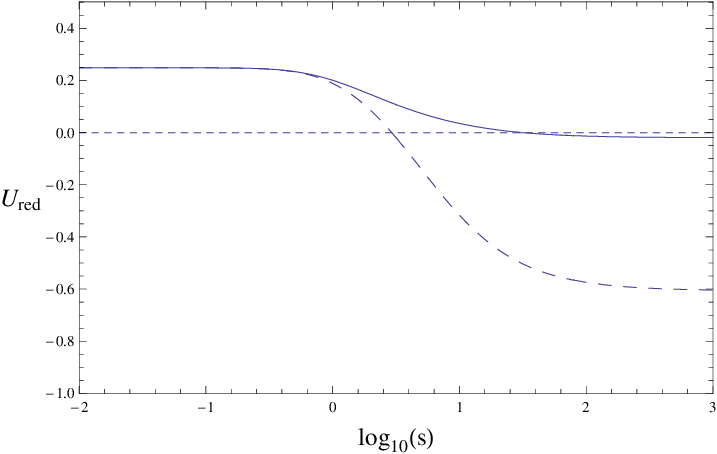}
   \put(-9.1,3.1){}
\put(-1.2,-.2){}
  \caption{}
\end{figure}
\newpage
\clearpage
\newpage
\setlength{\unitlength}{1cm}
\begin{figure}
 \includegraphics{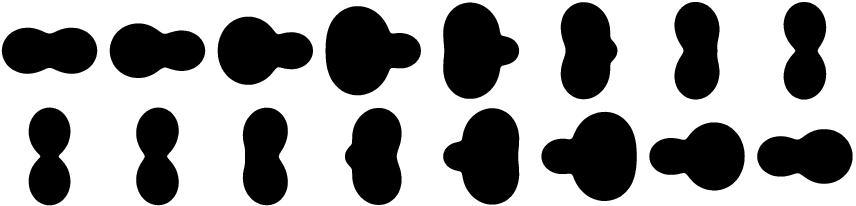}
   \put(-9.1,3.1){}
\put(-1.2,-.2){}
  \caption{}
\end{figure}
\newpage
\clearpage
\newpage
\setlength{\unitlength}{1cm}
\begin{figure}
 \includegraphics{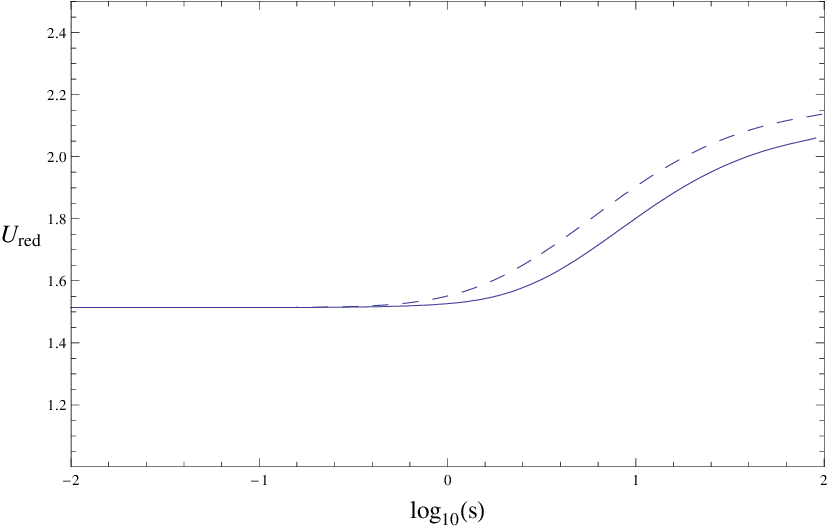}
   \put(-9.1,3.1){}
\put(-1.2,-.2){}
  \caption{}
\end{figure}
\newpage
\clearpage
\newpage
\setlength{\unitlength}{1cm}
\begin{figure}
 \includegraphics{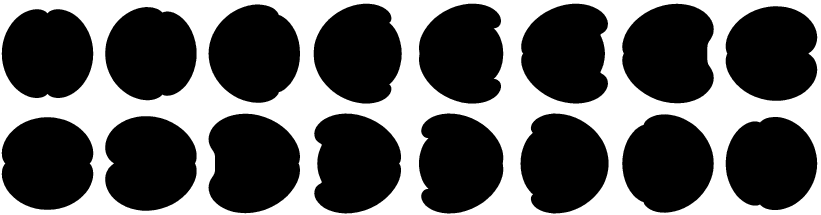}
   \put(-9.1,3.1){}
\put(-1.2,-.2){}
  \caption{}
\end{figure}
\newpage
\clearpage
\newpage
\setlength{\unitlength}{1cm}
\begin{figure}
 \includegraphics{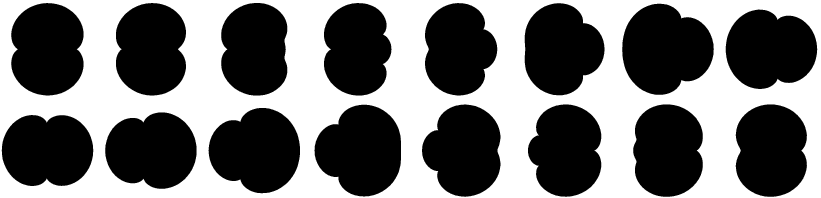}
   \put(-9.1,3.1){}
\put(-1.2,-.2){}
  \caption{}
\end{figure}
\newpage
\clearpage
\newpage
\setlength{\unitlength}{1cm}
\begin{figure}
 \includegraphics{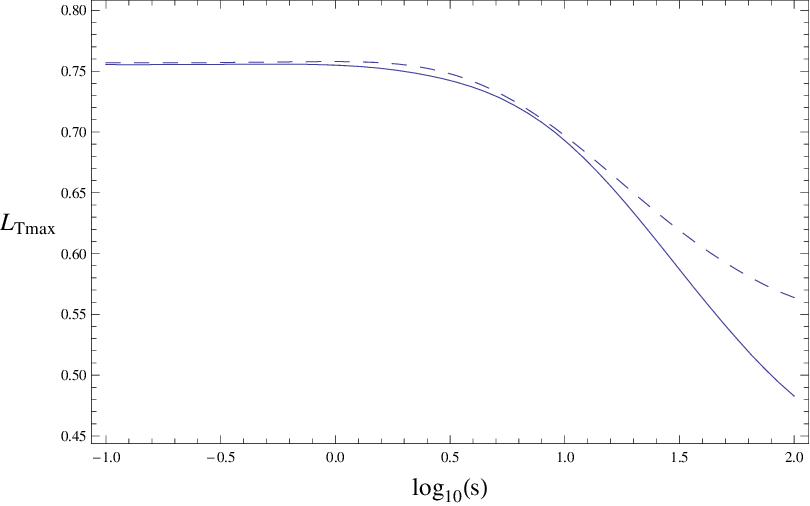}
   \put(-9.1,3.1){}
\put(-1.2,-.2){}
  \caption{}
\end{figure}
\newpage
\clearpage
\newpage
\setlength{\unitlength}{1cm}
\begin{figure}
 \includegraphics{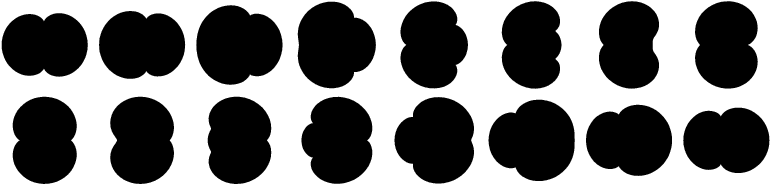}
   \put(-9.1,3.1){}
\put(-1.2,-.2){}
  \caption{}
\end{figure}
\newpage
\end{document}